\documentclass[fluids,article,accept,pdftex,moreauthors]{Definitions/mdpi}
\makeatletter
\let\c@lofdepth\relax
\let\c@lotdepth\relax
\makeatother

\usepackage{subfigure}
\makeatletter
\renewcommand{\@thesubfigure}{\normalsize(\textbf{\alph{subfigure}})}
\makeatother

\usepackage{epsfig}
\usepackage[percent]{overpic}

\def\bs{\begin{subequations}}
\def\es{\end{subequations}}
\def\be{\begin{equation}}
\def\ee{\end{equation}}
\def\bml{\begin{multline}}
\def\eml{\end{multline}}
\def\beq{\begin{eqnarray}}
\def\eeq{\end{eqnarray}}
\def\bd{\begin{description}}
\def\ed{\end{description}}
\def\bn{\begin{enumerate}}
\def\en{\end{enumerate}}
\def\bi{\begin{itemize}}
\def\ei{\end{itemize}}
\def\brem{\begin{remark}}
\def\erem{\end{remark}}
\def\bp{\begin{proof}}
\def\ep{\end{proof}}

\newcommand{\dd}[2]{\partial_{#1}{#2}}

\newcommand{\f}[2]{\frac{#1}{#2}}

\renewcommand{\vec}[1]{\mbox{\boldmath $ #1$}}

\newcommand{\B}{\vec B}

\newcommand{\ru}{\hat{\vec r}}

\newcommand{\kk}{\hat{\vec k}}
\newcommand{\n}{\nabla}

\newcommand{\x}{\times}
\newcommand{\rhobar}{\bar{\rho}}
\newcommand{\Tbar}{\bar{T}}
\newcommand{\Pbar}{\bar{P}}
\newcommand{\nubar}{\bar{\nu}}
\newcommand{\kapbar}{\bar{\kappa}}
\newcommand{\Nrho}{N_\rho}
\newcommand{\pol}{v}
\newcommand{\tor}{w}
\newcommand{\polB}{h}
\newcommand{\torB}{g}

\newcommand{\R}{\mathrm{R}}
\renewcommand{\Pr}{\mathrm{Pr}}
\newcommand{\Pm}{\mathrm{Pm}}

\usepackage{soul}

\allowdisplaybreaks
\usepackage{color}

\firstpage{1}
\pubvolume{8}
\issuenum{11}
\articlenumber{288}
\pubyear{2023}
\copyrightyear{2023}
\externaleditor{Academic Editors: D. Andrew S. Rees and Gang Li}
\datereceived{23 September 2023}
\daterevised{12 October 2023} 
\dateaccepted{24 October 2023}
\datepublished{27 October 2023 }
\hreflink{https://doi.org/10.3390/\linebreak fluids8110288} 

\Title{Differential Rotation in Convecting Spherical Shells with
Non-Uniform Viscosity and Entropy Diffusivity}
\TitleCitation{Differential Rotation in Convecting Spherical Shells with
Non-Uniform Viscosity and Entropy Diffusivity}

\Author{Parag Gupta 
\href{https://orcid.org/0000-0002-2976-5993}{\orcidicon}, David
MacTaggart *\href{https://orcid.org/0000-0003-2297-9312}{\orcidicon} and Radostin D. Simitev \href{https://orcid.org/0000-0002-2207-5789}{\orcidicon}}

\AuthorNames{Parag Gupta, David MacTaggart and Radostin  D. Simitev}
\AuthorCitation{Gupta, P.; MacTaggart, D.; Simitev, R.D.}
\address[1]{School of Mathematics and Statistics, University of Glasgow, Glasgow G12 8QQ, UK; {p.gupta.1@research.gla.ac.uk (P.G.); radostin.simitev@glasgow.ac.uk (R.D.S.)} 
\\}
\corres{\hangafter=1 \hangindent=1.05em \hspace{-0.82em}Correspondence: david.mactaggart@glasgow.ac.uk}

\abstract{
Contemporary three-dimensional physics-based simulations of the solar convection zone disagree with observations. They feature differential rotation substantially different from the true rotation inferred by solar helioseismology and exhibit a conveyor belt of convective ``Busse'' columns not found in observations. To help unravel this so-called ``convection conundrum'', we use a three-dimensional pseudospectral simulation code to investigate how radially non-uniform viscosity and entropy diffusivity affect differential rotation and convective flow patterns in density-stratified rotating spherical fluid shells. We find that radial non-uniformity in fluid properties enhances polar convection, which, in turn, induces non-negligible lateral entropy gradients that lead to large deviations from differential rotation geostrophy due to thermal wind balance. We report simulations wherein this mechanism maintains differential rotation patterns very similar to the true solar profile outside the tangent cylinder, although discrepancies remain at high latitudes. This is significant because differential rotation plays a key role in sustaining solar-like cyclic dipolar dynamos.}

\keyword{solar differential rotation; convection-driven spherical dynamo; magnetohydrodynamic numerical simulations}

\begin{document}
\section{Introduction}

Solar 
magnetic phenomena originate in the dynamic processes of thermal convection of electrically conducting plasma deep
within the solar interior.

High-resolution numerical simulations of thermal convection and magnetic field generation in rotating spherical shells
based on fundamental physical laws have emerged as useful tools for understanding the complexities of solar dynamics~\cite{Miesch2005,Charbonneau2020}.
Global simulations are increasingly compared to observational data and used to aid in observational interpretation. For instance, recent works have focused on leveraging time--distance helioseismology to constrain models of meridional circulation~\cite{Stejko2022}, while other studies have explored minimal
models to explain solar cycle dynamics by emphasizing the significance of non-axisymmetric ($m = 1$ or $m = 2$) components of the magnetic
field~\cite{simitev2012far}.

Despite the many advances made using numerical simulations, limitations and
challenges persist in capturing several fundamental aspects of the solar dynamo.
{The primary issue we investigate in this article is these challenges,} i.e., the so-called \emph{convective conundrum} 
\cite{OMara2016,Gizon2012}, which manifests as (a) a significant disparity between observed and simulated differential rotation and convective velocities and (b) the prominence of a
conveyor belt of large-scale convective columns (Busse columns) in simulations with little corresponding observational evidence
\cite{Charbonneau2020,hotta2023dynamics}.
Differential rotation, a crucial part of the Sun's global behaviour, is very well measured by helioseismic inversion~\cite{Thompson1996,Howe2009}. Its primary features are the variation in rotation speed with latitude, with the equator rotating faster (around 25~days) and the poles rotating slower (around 36 days), displaying a conical profile with a nearly radial orientation of angular velocity isocontours at medium latitudes.

{Simulations generally fail to produce isocontours of the angular velocity that are significantly inclined from the axis of rotation in the convection zone, and most simulations feature unrealistic isocontours parallel to the the axis of rotation instead
\cite{Warnecke2012,Guerrero2013,Matilsky2020}. Notable exceptions are the simulations reported in~\cite{Miesch2006},
where conically tilled contours were obtained by imposing a moderately large
latitudinal entropy gradient.} It is especially important to capture differential rotation correctly, as it plays a crucial role in magnetic field generation via the $\Omega$ effect of poloidal-to-toroidal field conversion, in addition to inducing dynamo wave oscillations akin to the solar cycle, as
demonstrated by flux-transport dynamo models~\cite{Dikpati1999} and some global simulations~\mbox{\cite{kapyla2012cyclic,Warnecke2018,simitev2012far,Simitev2015}}
.
{In the following sections, we demonstrate that a significantly large latitudinal entropy gradient can arise self-consistently without being additionally imposed, helping to improve agreement between our simulations \mbox{and observations}.}

In contrast to planetary cores and atmospheres, where basic density and material properties of the fluid change only weakly, in the solar interior, including the convection zone, there are very significant radial variations of thermodynamic quantities and properties.  Reference-state radial distributions of density, temperature and pressure in the convection zone can be estimated based on helioseismic arguments or solar evolution models (e.g.,~\cite{Brun2004}). In theory, molecular viscosity and thermal diffusivity profiles can be inferred from
molecular dynamics calculations (e.g.,~\cite{french2012ab}), but because
of the strong turbulence in the solar convection zone, it is unlikely
that these molecular values adequately represent the profiles of the effective viscosity and thermal diffusivity in global convective dynamo models. Hence, the radial distributions of viscosity and entropy diffusivity remain largely modelling choices---although rather important ones, as they are expected to influence the style and the spatial location of convection and, by extension, the properties of the global dynamo process. For example, the radial profile of entropy diffusivity directly affects
the entropy distribution, whose radial gradient consequently determines the local convective stability. Various choices of viscosity and entropy diffusivity profiles were investigated in the early study conducted by Glazmaier and Gilman~\cite{Glatzmaier1981}, who found that under a moderate shell rotation rate, convection moves from the outer to inner regions as the diffusivities are increased in the
outer regions and decreased in the inner regions. Several studies of solar convection have assumed that viscosity and diffusivity are functions involving a negative power of the mean density (e.g.,~\cite{Brun2002,Brun2004,Brun2022,Glatzmaier1981}) as opposed to the more commonly used uniform profiles~\cite{Warnecke2023,Matilsky2020,Warnecke2018,Karak2015,Simitev2015,Browning2010,Miesch2008}. In a recent study conducted by Sasaki et al.~\cite{Sasaki2018}, the effects of a radial distribution of entropy diffusivity on critical modes of
anelastic thermal convection in rotating spherical shell were studied.They found that convection morphology and location are strongly affected, but they restricted their analysis to the linear onset and considered only uniform viscosity. There is a need to extend the work reported in~\cite{Sasaki2018} to
non-linear regimes,

especially with an eye to improving the agreement between computed and observed solar differential rotation. This is the goal of the present study.

To this end, we report a set of numerical simulations based on the model proposed in~\cite{Simitev2015}, which we extended to incorporate radially non-uniform profiles of viscosity and entropy diffusivity. To pinpoint the effect of these assumptions, we performed computations and compared the results with those of baseline reference simulations in which uniform profiles were used. We consider both high and low values of the Prandtl number to make sure our findings are robust. The structure of this paper is outlined as follows. In Section~\ref{222}, we introduce the mathematical model used in this study. Section \ref{333} describes our main results with respect to the effects of non-uniform material properties on the differential rotation structure. In Section \ref{444}, 
we propose a likely mechanism to help explain our findings, provide evidence in the form of additional results and mention limitations of the study. We finish with a brief conclusion in Section 
\ref{555}.

\section{Mathematical Model}\label{222}

In this work, we extend the model of Jones at al.~\cite{Jones2011} by including radially non-uniform viscosity and entropy diffusivity. Our numerical
implementation, methods of solution and analysis follow
\cite{Simitev2015}. Details are recapitulated below for completeness.

We consider an electrically conducting perfect gas confined to a spherical shell. The shell rotates about the vertical axis with a fixed angular velocity ($\Omega_0 \mathbf{\kk}$), and an entropy contrast ($\Delta S$) is imposed between its inner and outer surfaces. Assuming a gravitational field  the inverse square of the radial distance from the centre ($r^{-2}$), a hydrostatic, polytropic reference state exists
\begin{gather}
\label{poly}
\rhobar = \rho_c\zeta^n, \quad \Tbar=T_c\zeta, \quad \Pbar = P_c
\zeta^{n+1},
\end{gather}
where $\zeta= c_0+c_1 d/r$ is a radial profile with parameters
$c_0=(2\zeta_o-\eta-1)/(1-\eta)$,
$c_1=(1+\eta)(1-\zeta_o)/(1-\eta)^2$ and $\zeta_o=(\eta+1)/(\eta
\exp(\Nrho/n)+1)$.
Constants $\rho_c$, $P_c$ and $T_c$ are reference values of density, pressure and temperature, respectively, in the middle of the shell. The gas polytropic index ($n$), the density scale height number
($N_\rho$) and the shell thickness ratio ($\eta$) are defined below.

Convection and magnetic field generation in this system are described by the  evolution equations of continuity, momentum, energy and magnetic flux. Using the Lantz--Braginsky formulation of the anelastic approximation
\cite{Jones2011}, these can be written in the following form: 
\vspace{-6pt}
\bs
\label{gov}
\begin{gather}
\label{gov.01}
\n \cdot \rhobar \mathbf{u} =0, \quad \n \cdot\B =0, \\
\label{gov.02}
\partial_t \mathbf{u} + (\n\x \mathbf{u})\x \mathbf{u}
=-\n\Pi -\tau(\kk\x \mathbf{u})+\f{\R}{\Pr}\f{S}{r^{2}}\ru
+
\frac{\rho_c}{\rhobar}\n \cdot  {\boldsymbol{\hat{\sigma}}}
+ \f{1}{\rhobar} (\n\x\B)\x\B, \\
\label{gov.03}
\partial_t S + \mathbf{u} \cdot \n S
= \f{1}{\Pr \rhobar\Tbar} \n \cdot\kapbar\rhobar\Tbar \n S
+ \f{c_1 \Pr}{\R \Tbar}\left(
{\boldsymbol{\hat{\sigma}}}: e
+ \f{1}{\Pm\rhobar}
(\n\x\B)^2
\right),
\\
\label{gov.04}
\partial_t \B = \n\x(\mathbf{u} \x\B)+\Pm^{-1} \n^2 \B,
\end{gather}
\es
where $\mathbf{u}$ is the velocity, $\B$ is the magnetic flux density, $S$
is the entropy, $\nabla \Pi$ includes all terms that can be written as gradients and 
$ {\boldsymbol{\hat{\sigma}}}$ is the deviatoric stress tensor:
\begin{gather}
\label{tens}
\sigma_{ij}=2\nubar\rhobar(e_{ij}-e_{kk}\delta_{ij}/3), \quad e_{ij}=(\partial_i u_j +\partial_j u_i)/2,
\nonumber
\end{gather}
where the double-dot symbol (\textbf{:}) denotes a component-wise inner product.

\textls[-15]{The governing equations are non-dimensionalized using the shell thickness \mbox{($d=r_o-r_i$)}} as a unit of length, $d^2/\nu_c$ as a unit of time, $\Delta S$ as a unit of entropy, $\nu_c \sqrt{\mu_0\rho_c}/d$ as a unit of magnetic induction, $\rho_c$ as a unit of density and $T_c$ as a unit of temperature. Here, $r_i$ and $r_o$ are the inner and the outer radius, respectively; $\lambda$ and $\mu_0$ are the magnetic diffusivity and permeability,
respectively; and $\nu$ and $\kappa$ are the viscosity and the
entropy diffusivity, respectively. The latter pair is assumed to have radially
non-uniform profiles:
\begin{gather}
\label{profiles}
\nu(r) = \nu_c \left(\frac{\overline{\rho}}{\rho_c}\right)^p, ~~~~ ~~~~\kappa(r) = \kappa_c \left(\frac{\overline{\rho}}{\rho_c}\right)^q,
\end{gather}
where $\nu_c$ and $\kappa_c$ are their values in the mid-shell, while $p$
and $q$ are real modelling constants.
The formulation is then characterized by seven non-dimensional parameters: the radius ratio, the polytropic index, the density scale number, the Rayleigh number, the thermal Prandtl number, the  magnetic Prandtl number and the Coriolis number, respectively defined as
\vspace{-6pt}
\begin{adjustwidth}{-\extralength}{0cm}
\begin{gather}
\label{params} 
\eta:=\frac{r_i}{r_o}, ~~~~
n, ~~~~
N_\rho:=\ln\frac{\rhobar(r_i)}{\rhobar(r_o)}, ~~~~
\R:=\frac{c_1 T_c d^2 \Delta S}{\nu_c  \kappa_c}, ~~~~
\Pr:=\frac{\nu_c}{\kappa_c}, ~~~~
\Pm:=\frac{\nu_c}{\lambda}, ~~~~
\tau := \frac{2\Omega_0 d^2}{\nu_c}.
\end{gather}
\end{adjustwidth}
Equations \eqref{gov.01}--\eqref{gov.04} are supplemented by no-slip boundary conditions on the
inner surface, with stress-free conditions on the outer surface
:
\bs
\label{BCS}
\begin{gather}
\pol = 0,
\quad
\partial_r \pol =0, \quad \tor = 0, \quad
\text{at}\quad r=r_i,\\
\pol = 0,
\quad
\partial_r^2 \pol - \f{\rhobar'}{\rhobar r}\dd{r}{} (r\pol) = 0,
\quad
\partial_r \tor - \f{\rhobar'}{\rhobar} \tor = 0 \quad
\text{at}\quad r=r_o.
\end{gather}
For
the entropy, a fixed contrast is imposed between the inner and outer surfaces:
\begin{gather}
S=1 ~~\text{ at }~~ r=r_i, \quad S=0 ~~\text{ at }~~ r=r_o.
\end{gather}
Lastly, ``vacuum'' boundary conditions for the magnetic field are derived from the assumption of an electrically insulating external region.
\begin{gather}
\torB =0, \quad  \polB-\polB^{(e)} = 0, \quad \partial_r
(\polB-\polB^{(e)})=0 \quad
\text{at}\quad r=r_i,r_o.
\end{gather}
\es
In the above,
$\pol$, $\tor$, $\polB$ and $\torB$ are poloidal and toroidal scalar fields in the following decompositions of of the momentum and magnetic fields, respectively:
\bs
\label{poltordef}
\begin{gather}
\rhobar \vec u = \nabla \times ( \nabla \times \ru r\pol) + \nabla \times
\ru r^2 \tor, \\
\vec B = \nabla \times  ( \nabla \times \ru \polB) + \nabla \times
\ru \torB,
\end{gather}
\es

The latter holds, since the mass flux ($\rhobar \mathbf{u}$) and the magnetic flux density ($\vec B$) are solenoidal vector fields. $\ru$ is the radial unit~vector.

\textls[-25]{The boundary value problem expressed by
Equations \eqref{gov.01}--\eqref{gov.04} and \mbox{conditions (5a)--(5d)}} 
 is {three-dimensional, time-dependent, highly coupled and non-linear; for these reasons, it can only be solved numerically. Our method of
numerical solution further exploits the poloidal--toroidal decomposition (6a) and (6b) as follows}. Equation~\eqref{gov.01} is satisfied
automatically. Upon the application of $\ru\cdot\n\x\n\x$ and $\ru\cdot\n\x$ to
Equation~\eqref{gov.02}, we obtain scalar equations for $v$ and $w$ and eliminate the pressure gradient. Similarly, applying \mbox{$\ru\cdot \n\x$ and $\ru\cdot$} to Equation~\eqref{gov.04}, we find equations for $h$ and $g$. The scalar unknowns ($\pol$, $\tor$, $\polB$, $\torB$ and $S$) are then expanded in Chebychev polynomials in the radial
variable ($r$) and in spherical harmonics in the angular variables $(\theta,\varphi)$. Their expansion coefficients are determined using a
pseudospectral method adapted from~\cite{Tilgner1999}. The calculations reported below were performed with a truncations of $71$, $193$ and $193$ for radial, zonal and latitudinal expansion coefficients, respectively
.

\section{Results}\label{333}

\subsection{Design of the Study and Choice of Parameter Values}

An evolution equation for the differential rotation generated by thermal convection in a rotating spherical shell can be obtained by averaging the zonal component of the momentum Equation \eqref{gov.02} over longitude. The result reveals that differential rotation is driven and maintained by a balance of Reynolds and Maxwell stresses modulated by stresses due to meridional circulation, magnetic tension and viscous dissipation~\cite{Miesch2005,Tassoul2000}. Naturally, parameter values affect all of these stresses. Aiming to isolate the effects of radially non-uniform viscosity and entropy diffusivity profiles on the overall shape and structure of the differential rotation, we keep as many of the non-dimensional parameters as possible at fixed values. In particular, we set
\begin{gather}
\label{parvals}
\eta=0.65, ~~~~ n = 2, ~~~~ N_\rho = 3, ~~~~ \R=3 \times 10^6, ~~~~ \tau = 2000.
\end{gather}
Fixing parameter values is also helpful in restricting the large eight-dimensional parameter space of the problem to a manageable size. At $\eta = 0.65$, the shell is slightly thicker than the solar convection zone ($\eta=0.7$), but this choice is made for ease of numerical simulation. The typical size of convective structures also depends on the thickness of the shell; thinner shells require spherical harmonic expansions of a higher order and degree to resolve the angular structure of flows. At $\tau = 2000$, the Coriolis number is moderately but not excessively large, reflecting the model assumption that the flow in the deep convection zone is buoyancy- rather than rotation-dominated, although the effects of rotation are also essential. The value of the polytropic index ($n = 2$) is larger than that for ideal gas ($n=3/2$), and the value of the density-scale height ($N_\rho = 3$) is much smaller than that estimated for the solar convection zone. However, increasing $N_\rho$ much beyond $5$ becomes computationally unfeasible. These choices are not expected to affect the dynamics significantly. They are also made for consistency and comparison with our earlier work~\cite{Simitev2015}.

The structure and intensity of the flow are most sensitive to the value of the Rayleigh number, which has been extensively studied (see,  e.g.,~\cite{Simitev2003,Simitev2015} and references therein). Here, we report a comparison of a few selected typical cases, which have an identical and moderately large value of the Rayleigh number. In Section \ref{4422}, we discuss an extended sequence of simulations in which the latter parameter is increased systematically from the onset. While the Rayleigh number dependence is relatively well understood, the dependence on the Prandtl number is much less so, although we refer to~\cite{SIMITEV2005,Kapyla2023}.

Linear stability analyses~\cite{Ardes1997,BUSSE2004}, as well as finite-amplitude simulations of thermal convection in a rapidly rotating thick shell geometry under Boussinesq approximation~\cite{Simitev2003,SIMITEV2005}, show that at significantly low Prandtl number values, convection develops in the so-called equatorially attached regime~\cite{Zhang1987,BUSSE2004} for which the Coriolis force does not play a strong role in the force balance. At Prandtl number values close to unity, the balance starts to include Coriolis effects, and the regime of spiraling convection is entered, where convective structures become elongated, protruding from the inner to the outer surface with strong spiralling. Thus, in order to capture, to some extent, the dependence on the Prandtl number, we contrast simulations conducted with small ($\Pr=0.3$), moderate ($\Pr=1$) and somewhat larger values ($\Pr=5$).

\begin{figure}[H]
\hspace*{-10mm}

\subfigure[]{\includegraphics[width=0.516\linewidth]{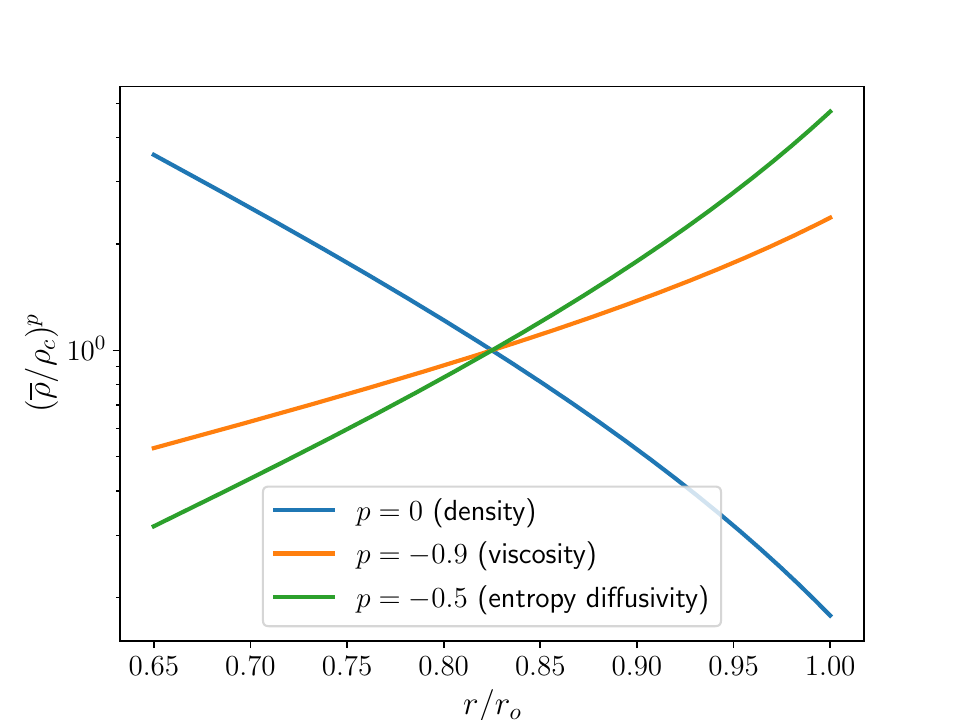}}\subfigure[]{\includegraphics[width=0.489\linewidth]{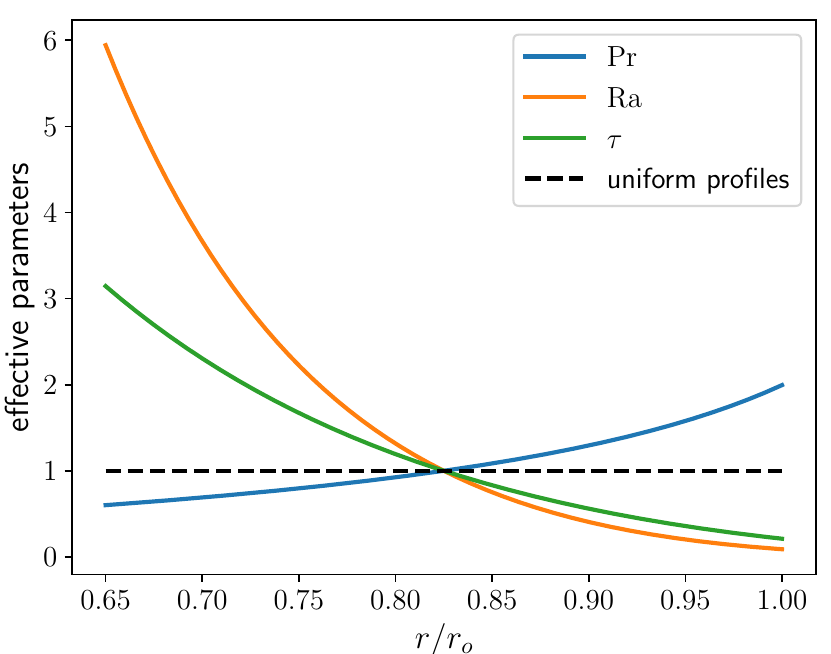}}

\caption{
\label{nfig010}
Radial 
profiles for different model parameters. (\textbf{a}) Non-uniform viscosity and entropy diffusivity vary relative to the density. (\textbf{b}) Non-dimensional parameters $\R$, $\Pr$ and $\tau$ vary when uniform and non-uniform viscosity and entropy diffusivity are considered. For display purposes, all parameters were normalized with respect to their values at the centre of the spherical shell. This allows for the visualization of how the parameters vary with respect to the uniform profile.}
\end{figure}

To address the objectives of this study, we then compare a set of simulations performed using uniform radial profiles of viscosity and entropy diffusivity to a set of simulations in which these two profiles vary in the radial direction as $\bar \nu \propto \rho^{-0.9}$ and $\bar\kappa\propto\rho^{-0.5}$, respectively. In the absence of stringent observational or accurate theoretical constraints, these dependencies were selected so as to maximize the deviation of the two profiles from the uniform profiles at the fixed parameter values mentioned above
and to ensure that well-resolved solutions are obtained. The two profiles, along with the radial profile of density in the model, are illustrated in Figure \ref{nfig010}a.

The values of the Prandtl, Rayleigh and Coriolis numbers discussed above correspond to the fluid properties in the middle of the spherical shell (see definition \eqref{params}
). Because of the non-uniform viscosity and the entropy diffusivity used in this study, the effective values of these quantities also vary in terms of radius, as illustrated in Figure \ref{nfig010}b
. This variation significantly affects the local morphology and intensity of the flows, as discussed further~below.

In summary of this discussion, in Table \ref{tab01}, we list the six main dynamo solutions on which we focus our attention for most of this article.

The table also provides the values of some of their most most important global average characteristics, including energy density components that characterize the various components of the flow. The mean and fluctuating toroidal and poloidal components of the total kinetic energy ($E_\text{tot}$) are defined~as
\vspace{-12pt}
\bs
\label{engs}
\begin{eqnarray}
\label{7}
\bar E_p = \langle \big(\nabla \times ( \nabla \bar \pol \times \vec r )
\big)^2/(2\rhobar)  \rangle,
&
\bar E_t = \langle \big(\nabla r \bar\tor \times \vec r \big)^2
/(2\rhobar)  \rangle, \\
\label{8}
\check E_p =  \langle \big(\nabla \times ( \nabla \check \pol
\times \vec r)  \big)^2/(2\rhobar) \rangle, &
\check E_t = \langle \big( \nabla r \check \tor \times\vec r \big)^2
/(2\rhobar) \rangle,
\end{eqnarray}
\es
where angular brackets denote averages over the spherical volume of the shell, bars denote axisymmetric parts and check marks denote non-axisymmetric parts of a
scalar field. The total magnetic energy ($M_\text{tot}$) can be split in a similar way, with components defined as in Equations (8a) and (8b) but with $\polB$ and $\torB$ replacing $\pol$ and $\tor$, respectively, and without the $\rhobar^{-1}$ factor within the angular brackets. The total energies are, of course, the sum of all components.
\vspace{-12pt}

\begin{table}[H]
\caption{\label{tab01} Summary 
of model parameter values and computed energy densities for the six selected dynamo solutions discussed in the text.}

\begin{adjustwidth}{-\extralength}{0cm}

\newcolumntype{C}{>{\centering\arraybackslash}X}
\begin{tabularx}{\fulllength}{CCCCCCC}
\toprule
\multirow{1}{*}{} & \multirow{1}{*}{\textbf{A}} & \multirow{1}{*}{\textbf{B}} & \multirow{1}{*}{\textbf{C}} & \multirow{1}{*}{\textbf{D}} & \multirow{1}{*}{\textbf{E}} & \multirow{1}{*}{\textbf{F}} \\
\midrule

\boldmath{$\mathrm{Type}$} &  \textbf{Uniform} & \textbf{Non-Uniform}  &  \textbf{Uniform} & \textbf{Non-Uniform} &  \textbf{Uniform} & \textbf{Non-Uniform} \\ \midrule

$\eta$ & 0.65 & 0.65  & 0.65 & 0.65  & 0.65 & 0.65 \\
$R$ & $3 \times 10^6$ & $3 \times 10^6$  & $3 \times 10^6$  & $3 \times 10^6$ & $3 \times 10^6$ & $3 \times 10^6$\\
$\tau$ & 2000  & 2000  &  2000  &  2000 &  2000 &  2000\\
$P_r$ &  0.3 &  0.3 & 1 & 1 & 5 & 5    \\
$N_\rho$  &  3 & 3  & 3  & 3  & 3 & 3\\
$n$ &  2 & 2  & 2  & 2  & 2  & 2 \\
$P_m$ &  1  &  1  & 4 & 4  & 10 & 10  \\
\midrule
$\overline{E}_{tot} $
& 11,785.6
& 18,627.4
& 846.488
& 1006.96
& 29.1356
& 20.9883\\
$\overline{E}_p $
& 23.7465
& 64.0577
& 0.97983
& 4.29216
& 0.0215626
& 0.0870345\\
$\overline{E}_t $
& 5930.15
& 11,540.3
& 182.149
& 570.75
& 4.7595
& 1.32042\\
$\check{E}_p   $
& 1946.24
& 1746.21
& 271.194
& 175.511
& 11.0902
& 6.81947\\
$\check{E}_t $
& 1574.74
& 2261.39
& 207.107
& 218.885
& 9.35735
& 11.5854\\
\midrule
$\overline{M}_{tot} $
& 1189.03
& 0.0662693
& 464.484
& 14.92
& 21.4386
& 0.638192\\
$\overline{M}_p $
& 6.26582
& 0.00139628
& 4.48705
& 0.117454
& 0.463097
& 0.0718596\\
$\overline{M}_t $
& 13.3781
& 0.0012872
& 6.38913
& 0.129885
& 0.43025
& 0.0194026\\
$\check{M}_p    $
& 256.771
& 0.0164442
& 111.957
& 4.44414
& 5.34042
& 0.181853\\
$\check{M}_t $
& 290.496
& 0.0153525
& 105.607
& 3.82462
& 4.41429
& 0.0936899 \\
\midrule
$Rm $
& 153.5291
& 193.0150
& 164.5828
& 44.8767
&  76.3350
& 64.7890\\
$Ro$
& 0.1535
& 0.1930
& 0.0411
& 0.0448
& 0.0076
& 0.0064\\
$Lo$
& 0.0487
& 0.0003640
& 0.0304
& 0.0054
& 0.0065
& 0.001129\\
\bottomrule
\end{tabularx}
\end{adjustwidth}
\end{table}

{The parameter values reported in Table \ref{tab01}, along with the values for the solar mass ($M_\odot = 1.98\times10^{30}~\text{kg}$), the
Carrington sidereal
rotation ($\Omega_0=\Omega_\odot = 2.87 \times 10^{-6} ~\text{rad s}^{-1}$),
the outer solar radius ($r_o=0.95 r_\odot = 6.59 \times 10^7
~\text{m}$), the gravitational  constant \linebreak \mbox{($G=6.67 \times 10^{-11}~\text{m}^3 \text{kg}^{-1}
\text{s}^{-2}$)}, and a solar central density \textls[-35]{estimate ($\rhobar_c= 1.622 \times 10^5
~\text{kg/m}^3$)}, can be used with polytropic profile \eqref{poly} and definition \eqref{params} 
to estimate transport coefficients such as the viscosity, entropy and magnetic diffusivities in dimensional form, as well as derived thermodynamic quantities, such as polytropic constants and luminosities for each of the simulations. Unsurprisingly, comparison with known solar plasma values 
reveals differences, which are justified in the preceding discussion. The values we use must be interpreted as effective ``eddy'' coefficients that---at least in part---account for unresolved turbulent subgrid scales.}

\subsection{Direct Comparison between Simulated Differential Rotation and Observations}

\begin{figure}[H]

\includegraphics[width=0.99\linewidth]{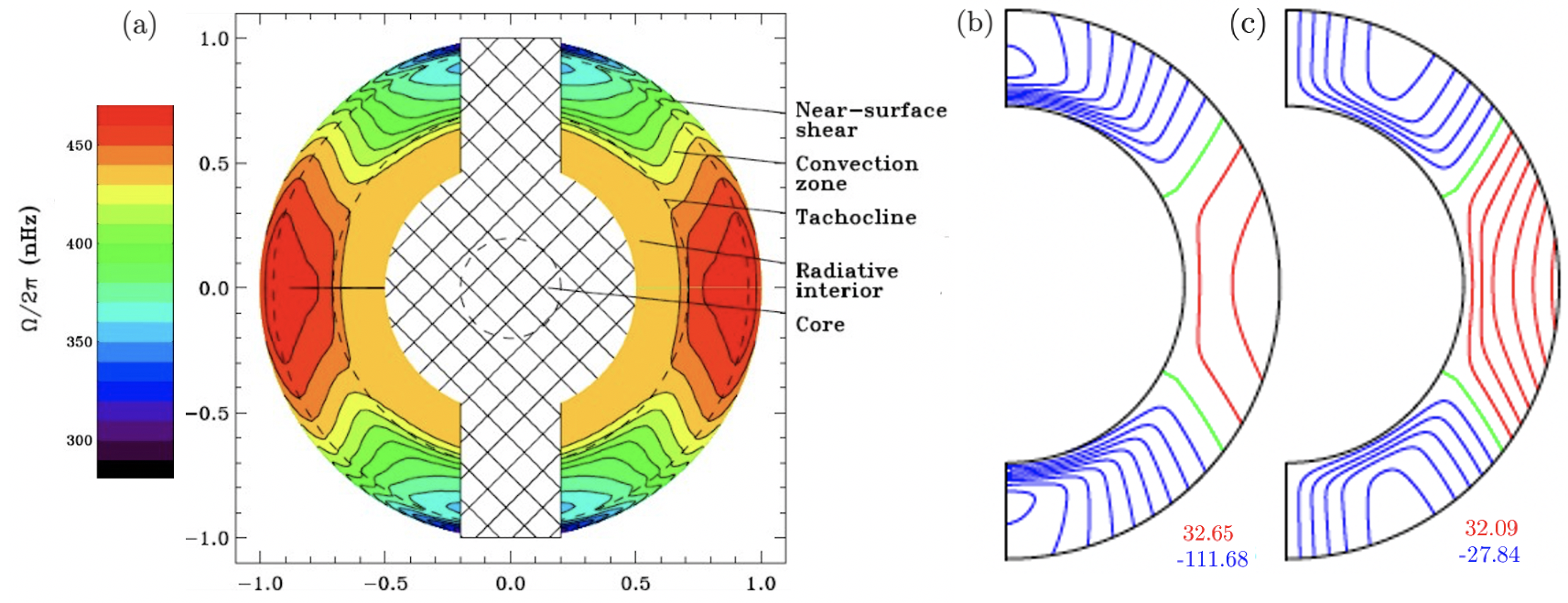}

\caption{
\label{nfig020}
The 
observed profile of the solar differential rotation.
(\textbf{a}) Cross-sectional view of the Sun's interior, depicting contours of constant angular velocity ($\Omega$) temporally averaged over 12 years of Stanford Michelson Doppler Imager (MDI) data.  Image adapted from Howe~\cite{Howe2009} (Springer Nature, licensed by CC BY 4)
.
(\textbf{b}) The solar angular velocity relative to the rotation frame of the Sun $(\Omega -\Omega_\odot)$, with solar frame rotation of $\Omega_\odot = 870\pi$ nHz. A closed-form approximation is used according to Kosovichev~\cite{Kosovichev1996}
. (\textbf{c}) Azimuthally averaged zonal velocity ($r \sin \theta (\Omega -\Omega_\odot)$) in the rotating frame is used for comparison with simulation results. Maximum and minimum values are indicated in (\textbf{b},\textbf{c}), respectively; contour lines are equidistant, with positive levels in red and negative levels in blue (this style is used throughout).}
\end{figure}

To assess the effects of non-uniform viscosity and entropy diffusivity profiles on the overall shape and structure of the differential rotation, it is desirable to compare the solutions of the model to the observed true profile of the solar differential rotation. The latter is well measured~\cite{Thompson1996}, as previously mentioned, and the angular velocity ($\Omega$) is shown in Figure \ref{nfig020}a courtesy of Howe~\cite{Howe2009}. The strongest rotation is concentrated near the equator, with contours of constant rotation closely following lines of oriented 
25$^\circ$ to the axis of rotation. It is these feature that give rise to what we refer to as ``conical'' differential rotation (as opposed to ``cylindrical'' rotation parallel to the axis of rotation
). To aid in our numerical implementation, we chose to work with a simpler but, for our purposes, sufficiently accurate analytical approximation constructed by Kosovichev~\cite{Kosovichev1996}. This is illustrated in Figure \ref{nfig020}b, where we omitted the quiescent radiative interior, the thin solar near-surface shear layer and the tachocline layer, as these are not captured by our mathematical model. Since Equations \eqref{gov.01}--\eqref{gov.04} of our model are formulated with respect to a rotating coordinate system, we subtracted the rigid frame rotation of the Sun ($\Omega_\odot = 870 \pi$ nHz). Figure \ref{nfig020}c shows the corresponding zonal (linear) velocity ($u_\varphi = r \sin \theta (\Omega-\Omega_\odot)$, where the distance from the centre ($r$) is measured in the dimensional units of our model (shell thickness)).

\begin{figure}[H]
\epsfig{file=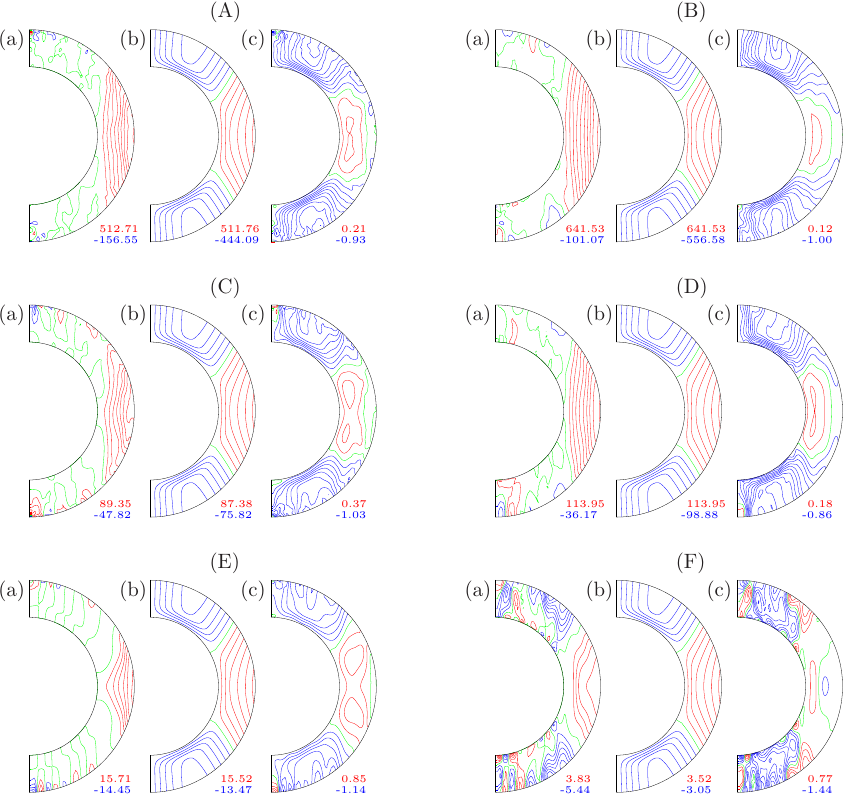,width=0.99\linewidth,clip=}
\caption{\label{nfig030} Differential 
rotation of the dynamo solutions under consideration. The three rows correspond to (\textbf{A},\textbf{B}) $\Pr=0.3$, (\textbf{C},\textbf{D}) $\Pr=1$ and (\textbf{E},\textbf{F}) $\Pr=5$; other parameters have the same values as in \eqref{parvals}
. Cases (\textbf{A},\textbf{C},\textbf{E}) have uniform viscosity and entropy diffusivity profiles; cases (\textbf{B},\textbf{D},\textbf{F}) have non-uniform profiles. In each group of three subfigures, (\textbf{a}) displays isocontours of the azimuthally averaged zonal velocity ($\overline{u}_\varphi$), (\textbf{b}) displays the reference observational profile of differential rotation ($r \sin \theta (\Omega - \Omega_\odot$)) rescaled to obtain the same maximum at the equator as in (\textbf{a},\textbf{c}) displays the relative difference between (\textbf{a},\textbf{b}) relative to the maximum of (\textbf{b}).}
\end{figure}

Figure \ref{nfig030} shows a direct comparison of the azimuthally averaged zonal velocity ($\bar u_\varphi$) of our model solutions with the latter profile (Figure \ref{nfig020}c). Our primary interest here is the shape of the differential rotation profile, so we rescaled the reference profile to obtain the same maximum at the equator as the simulation results. We scaled with respect to the equatorial maximum, since our model is primarily concerned with the profile in the tangent~cylinder.

\textls[-5]{For the smallest value of the Prandtl number ($\Pr=0.3$) and with uniform profiles of $\nubar$ and $\kapbar$, Figure \ref{nfig030}(Aa) shows that the differential rotation is geostrophic, with contour lines of $\bar u_\varphi$ parallel to the rotation axis and a prominent prograde jet filling the region outside of the tangent cylinder (the coaxial cylinder 
with frame rotation axis and tangential to the inner core of the shell at the equatorial plane), while within the tangent cylinder, the zonal velocity vanishes. This pattern compares poorly with the solar profile shown in Figure~\ref{nfig030}(Ab), which is also reflected in Figure \ref{nfig030}(Ac), where errors appear across all latitudes. The inclusion of radially non-uniform profiles of $\nubar$ and $\kapbar$ seems to have a negligible effect on the differential rotation, which retains the structure just described (Figure \ref{nfig030}B). Somewhat surprisingly, little further change to the pattern is seen at the moderate value of the Prandtl number ($\Pr=1$) in both uniform and non-uniform profiles  of $\nubar$ and $\kapbar$ (Figure \ref{nfig030}C,D). At a large value of the Prandtl number ($\Pr=5$), the contours of zonal velocity start to deviate from a cylindrical shape (Figure \ref{nfig030}(Ea)). The effect is strongly amplified in the case of non-uniform viscosity and entropy diffusivity (Figure \ref{nfig030}(Fa)) and assumes a shape very similar to that of the observed solar rotation profile shown in Figure \ref{nfig030}(Fb), resulting in vanishingly small discrepancies at mid latitudes (Figure \ref{nfig030}(Fc)). It is notable that in the latter case, differential rotation develops within the tangent cylinder at high latitudes and in the polar regions. While the structure of the polar differential rotation is more complicated than the observed rotation, it is notable that it features a relatively large retrograde polar jet, as does the  true solar zonal flow. We note that comparison in the polar regions is subject to greater uncertainty. Firstly, in simulations, numerical error increases when the distance from the axis of rotation tends toward zero, which occurs in the polar region. Secondly, the polar regions of the Sun are not in a direct line of sight, so observational measurements in these regions are less accurate.}

\begin{figure}[H]

\begin{adjustwidth}{-\extralength}{0cm}
\centering 
\epsfig{file=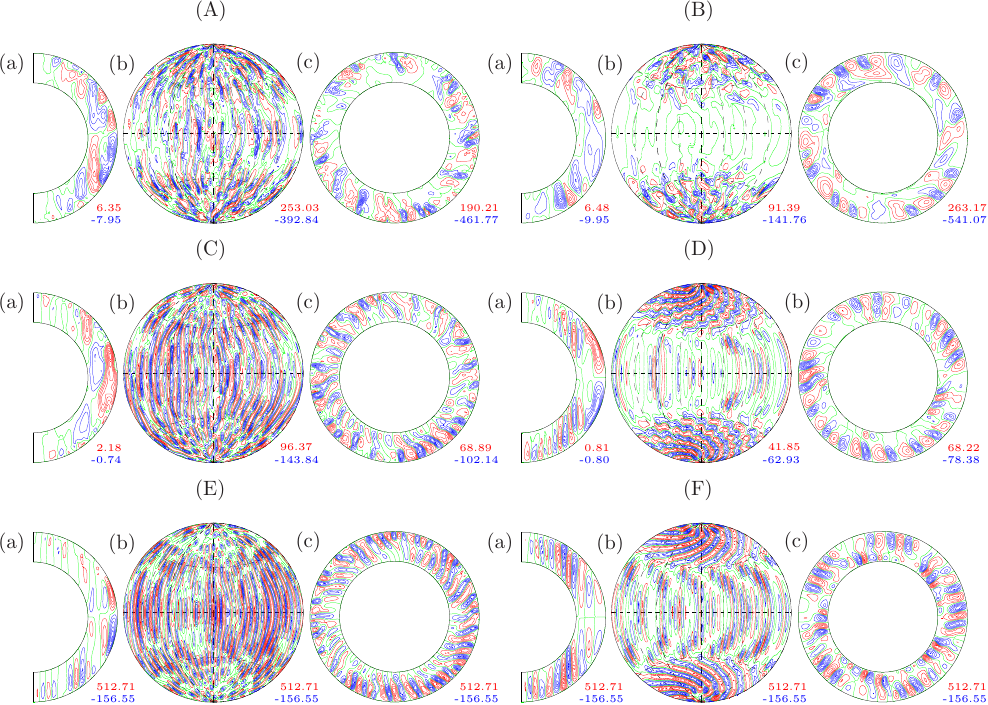,width=0.9\linewidth,clip=}\\
\end{adjustwidth}
\caption{\label{nfig040}
Flow 
structures of the cases shown in Figure \ref{nfig030} and labeled in the same way. In each group, (\textbf{a})~displays azimuthally averaged meridional circulation in a meridional plane, (\textbf{b}) displays isocontours of the radial velocity ($u_r$) on the spherical surface at $r=0.5$ and (\textbf{c}) shows poloidal streamlines of the velocity field in the equatorial plane.}
\end{figure}

\begin{figure}[H]

\begin{adjustwidth}{-\extralength}{0cm}
\centering 
\epsfig{file=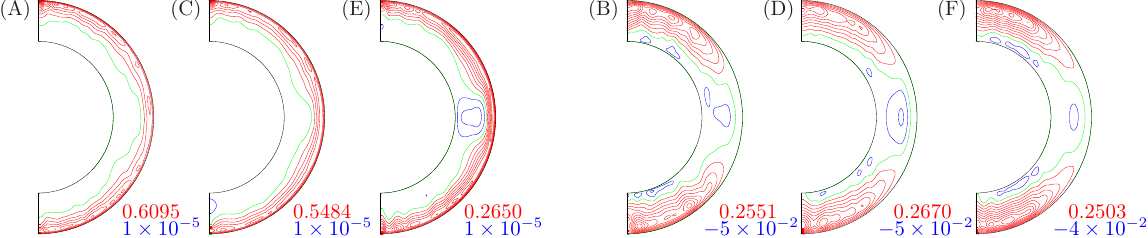,width=0.95\linewidth,clip=}
\end{adjustwidth}
\caption{\label{nfig050}
\textls[-15]{Azimuthally 
and time-averaged entropy ($\langle S\rangle_{\varphi,t}$) for the cases plotted in Figures \ref{nfig030} and \ref{nfig040} and labeled in the same way, i.e., uniform profiles (\textbf{A},\textbf{C},\textbf{E}) and non-uniform profiles (\textbf{B},\textbf{D},\textbf{F}) of viscosity and entropy diffusivity, as well as (\textbf{A},\textbf{B}) $\Pr=0.3$, (\textbf{C},\textbf{D}) $\Pr=1$ and (\textbf{E},\textbf{F}) $\Pr=5$, with other parameters from by~\eqref{parvals}
. }}
\end{figure}

{In summary, the simulation that produces the most solar-like differential rotation is that described in column F of Table \ref{tab01}. Details of this simulation are displayed in \mbox{Figures
\ref{nfig030}F, \ref{nfig040}F and \ref{nfig050}F} and discussed further below.}

\vspace{-3pt}
\unskip

\section{Discussion Including Further Results}\label{444}

\subsection{Structure of the Flow and Thermal Wind Balance}

In order to understand the increasing agreement between observed and measured zonal flows in the presence of non-uniform  viscosity and entropy diffusivity at larger values of the Prandtl number, in Figure \ref{nfig040}, we visualize the rest of the fluid flow components for the same simulations discussed in relation to Figure \ref{nfig030}. Outside the tangent cylinder, convection occurs in the form of rolls aligned with the axis of rotation. These exhibit properties of thermal
Rossby waves in that they drift in the prograde azimuthal direction. Since the variation in the direction of the axis of
rotation is minimized by this form of convection, it is best visualized by the streamlines in the equatorial plane, as shown in the third and sixth columns (c) of Figure \ref{nfig040}. These equatorial cross sections indicate the spiralling shape of the convection columns, which becomes less pronounced with the increase in the Prandtl number ($\Pr$). The
convection columns extend to fill the entire tangent cylinder, as
visible in the plots of the radial velocity projected on the spherical
surface at mid-shell (second and fifth columns (b) of Figure
\ref{nfig040}). The most striking effect of the presence of non-uniform profiles of $\nubar$ and $\kapbar$ in comparison with the corresponding uniform-profile simulations is the development of dominant convection flow in the polar regions inside the tangent cylinder. As gravity and rotation vectors are nearly parallel in the polar regions, convection resembles that realized in a horizontal layer heated from below and rotated about a vertical axis but now modulated by interactions with the convection outside of the tangent cylinder. As noted in relation to Figure \ref{nfig030}, differential rotation in the polar regions is retrograde, therefore tending to reduce the rotational constraint and forming a feedback loop, supporting the polar flow. At the largest value of the Prandtl number ($\Pr=5$), the polar flows become particularly strong and organized into elongated rolls in the shape of bicycle wheel spokes emanating from the axis of rotation and gently spiraling in the retrograde direction towards the periphery opposite to the drift of the equatorial structures.

The presence of relatively vigorous and well-organized polar convection suggests that entropy transport is enhanced in the polar regions compared to that in the equatorial regions. To verify this, in Figure \ref{nfig050}, we plot the gradient of entropy with respect to latitude for all cases discussed in Figures \ref{nfig030} and  \ref{nfig040}.

There is a clear distinction between the uniform cases (Figure \ref{nfig050}A,C,E) and the non-uniform cases (Figure \ref{nfig050}B,D,F), with a break in positive $\langle S\rangle_{\varphi,t}$ from mid to high/low latitudes. This result is significant because deviations of differential rotations form geostrophy (cylindrical profiles) are known to also be enhanced by the presence of non-vanishing latitudinal entropy gradients~\cite{Tassoul2000,Pedlosky1987,Miesch2005}. Indeed, when pressure gradients and Coriolis and buoyancy forces are in a
dominant balance, as occured in our simulations and may indeed occur in
the bulk of the solar convection zone, it can be shown that the zonal
component of the curl of the momentum Equation \eqref{gov.02} reduces to the so-called thermal wind balance
\begin{gather}
\label{eq:thermalwind}
\kk \cdot \nabla \langle \mathbf{u}_\varphi \rangle_t ~~\propto~~ \frac{\partial \langle
S\rangle_{\varphi,t}}{\partial\theta},
\end{gather}
which is a generalisation of the Taylor--Proudman theorem for rotating fluids in the presence of buoyancy. This relation shows that in or close to an adiabatic state, if \mbox{${\partial \langle
S\rangle_{\varphi,t}}/{\partial\theta} \approx 0$}, then the rotation profile must be close to cylindrical and when, on the other hand, significant latitudinal entropy gradients are present, as in the cases of non-uniform viscosity and diffusivity profiles shown in Figure \ref{nfig050}, where non-cylindrical differential rotation \mbox{is promoted}.

{To summarise, the radially non-uniform viscosity and entropy diffusivity profiles allow enhanced convection to develop at the poles, resulting in a non-vanishing entropy gradient with increased latitude
. In turn, this helps to produce more solar-like differential rotation according to Equation \eqref{eq:thermalwind}.
}

\subsection{Secondary Considerations}\label{4422}

In this preliminary study, we did not fully explore the behaviour model solutions in the accessible parameter space, as we wished to focus our attention on the main new extensions of our model, i.e., the introduction of non-uniform $\nubar$ and $\kapbar$ profiles. We are, however, aware of other effects and processes that may alter the results described above, at least quantitatively. We briefly touch upon two of them now.

\begin{figure}[H]
\epsfig{file=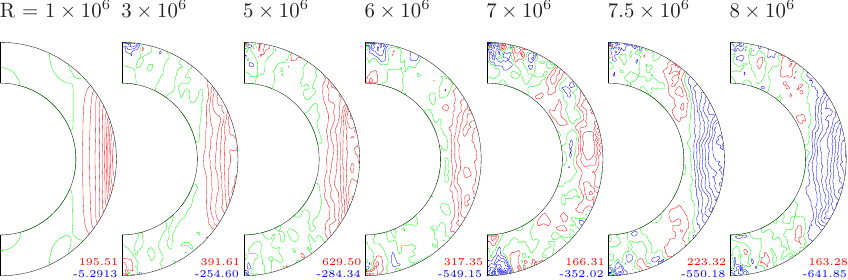,width=0.99\linewidth,clip=}\\
\caption{\label{nfig060}
Differential 
rotation as a function of the Rayleigh number and the solar/antisolar transition. Isocontours of azimuthally averaged zonal velocity ($\overline{u}_\varphi$) are plotted for the Rayleigh number values indicated in the plot. The rest of the parameter values are specified in \eqref{parvals}, with $\Pr=0.3$ and uniform $\nubar$ and $\kapbar$ values.}
\end{figure}

\emph{Solar--antisolar transition.}
Firstly, in Figures \ref{nfig030}--\ref{nfig050} we restricted the discussion to a few cases with an identical but fixed value/profile of the Rayleigh number. The Rayleigh number measures the contribution of buoyancy to the overall balance of forces in the system; even in the presence of uniform
viscosity and entropy diffusivity, it has a profound effect on convection. This dependence has been studied extensively
using linear instability analysis, as well as finite-amplitude simulations under various conditions and approximations. Some
references to our own works are~\cite{Simitev2003,Busse2005,Mather2020}, and recent review papers include~\cite{Charbonneau2020}. Here, we only remark that with the increase in driving due to buoyancy, an abrupt transition occurs from prograde solar-like differential rotation to retrograde antisolar differential rotation in the equatorial region. Figure \ref{nfig060}	illustrates the phenomenon for a set of cases with increasing Rayleigh number. This transition has been known since the early works reported in~\cite{Gilman1979} and has recently attracted attention from a number of authors in the context of solar and stellar convection~\cite{Simitev2015,Kapyla2023}. Since
buoyancy is expected to predominate in the force balance of solar convection, the results reported above for a moderate value of the Rayleigh number may now appear transient. However, the critical value of the Rayleigh number ($\R$) for transition from solar to antisolar rotation is a function of the remaining parameter values. For instance, while we were able to observe a transition at $\Pr=0.3$, as shown in Figure~\ref{nfig060}, no such transition was observed at $\Pr=1$ and $\Pr=5$
. The transition depends,
for instance, on  the value of the background density stratification parameter
($N_\rho$), which, in our simulations, was significantly far removed from estimates for the solar convection zone. Hence, we hypothesize that in the regimes of prograde equatorial rotation, the effects
reported in our work still hold true. Another observation to note in Figure \ref{nfig060} is that the strong buoyancy driving convection towards turbulent states is not sufficient to produce conical differential rotation on its own; in fact, before and after, the transition rotation remains largely geostrophic in the illustrated uniform profile cases.

\begin{figure}[H]

\includegraphics[width=0.998\linewidth,clip=on]{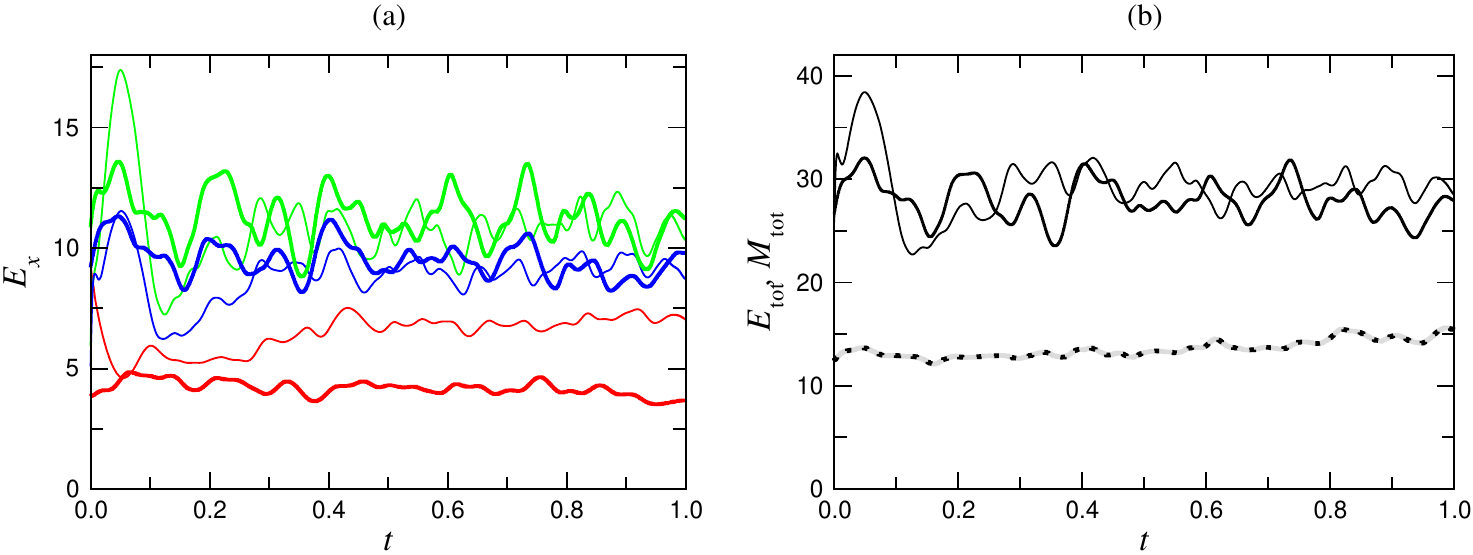}

\caption{\label{nfig070}
\textls[-25]{{Time} 
series of energy densities of the self-sustained dynamo case first shown in Figure~\ref{nfig030}E (thick lines) against non-magnetic convection under identical parameters (thin lines labeled E$'$ in \mbox{Figure \ref{nfig080}}). (\textbf{a})~Selected kinetic energy densities. The equatorially symmetric components of the mean toroidal ($\bar E_t^s$ (red)), fluctuating poloidal  ($\check E_p^s$ (green)) and fluctuating toroidal  ($\check E_t^s$ (blue)) energies. (\textbf{b}) Total kinetic energy densities and total magnetic energy density for the dynamo case (grey dotted line).} }
\end{figure}

\begin{figure}[H]
\epsfig{file=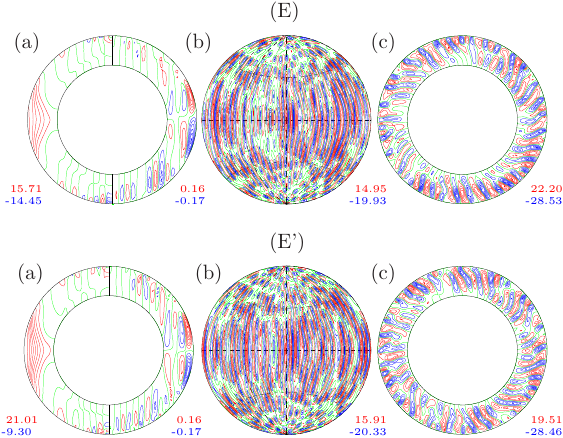,width=0.9\linewidth,clip=}
\caption{
\label{nfig080}
Comparison 
of a dynamo solution (E) and a non-magnetic convection solution (E$'$) under identical parameter values. The dynamo is the same case (E) plotted in Figures \ref{nfig030}--\ref{nfig050} and \ref{nfig070}. The contour plots in (\textbf{a}--\textbf{c}) show the same solution components as those plotted in (\textbf{a}--\textbf{c}) of Figure \ref{nfig040}, plus the differential rotation in the left half of (\textbf{a}).}
\end{figure}

\begin{figure}[H]
\epsfig{file=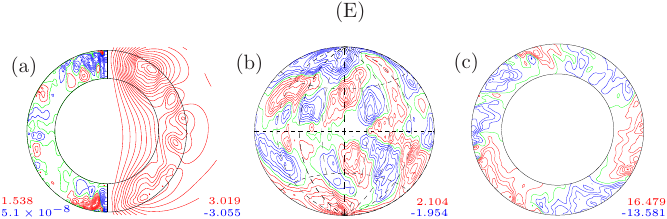,height=0.33\linewidth,width=0.998\linewidth,clip=}
\caption{
\label{nfig090}
Magnetic 
field components of the dynamo solution (E) of Figure \ref{nfig080}. (\textbf{a}) Toroidal and poloidal field lines in the meridional plane to the left and right, respectively. (\textbf{b}) Radial magnetic field continued slightly above the shell surface ($B_r(r=1.2)$) (\textbf{c}) Field lines in a plane parallel to the equator but at a latitude of $\theta=30^\circ$.
}
\end{figure}

\emph{Effects of self-sustained magnetic fields.}
So far, we have omitted from the discussion the effects of self-sustained magnetic fields on the convective flow and the associated differential rotation. Figures \ref{nfig070} and \ref{nfig080} show a comparison of a non-magnetic convection simulation with a dynamo solution obtained with identical parameter
values and with a magnetic Prandtl number of $\Pm=10$. The value of the magnetic Prandtl number was selected so as to ensure that a non-decaying magnetic field was obtained. Figure \ref{nfig070}, in particular, shows a direct comparison of
the kinetic energy components of the two cases. The main effect of the self-generated and self-sustained magnetic field in the dynamo simulation is a reduction of the kinetic energy component associated with the zonal \mbox{flow---by} roughly {25\%} in this case. This is  due to the fact that even in this significantly diffusive case, the poloidal
components of the magnetic field behave, to some extent, as frozen in the fluid and thus act to oppose and slow down zonal jets. Zonal flow, by its nature, does not convect heat between the spherical boundaries but distorts and suppresses  convective motions. Thus, because
the magnetic field now acts to reduce zonal flow, the mean and fluctuating poloidal and toroidal kinetic energies are somewhat larger but not to a significant degree. \mbox{Figure \ref{nfig070}} further
demonstrates the chaotic temporal behaviour of the solutions. \mbox{Figure \ref{nfig080}} shows a side-by-side comparison of the spatial structure of the non-magnetic and dynamo solutions in this case. Apart from a small difference in differential rotation isocontours, which are slightly more ``pinched'' in the non-magnetic case, the rest
of the flow components appear nearly indistinguishable, including
dominant azimuthal, radial and latitudinal wave numbers and typical length scales. Therefore, we must conclude that the self-sustained magnetic field does not substantially alter the convective and zonal flows. Additionally, \mbox{Figure \ref{nfig090}} illustrates the morphology of the generated magnetic field. The dynamo exhibits a dominant bipolar symmetry in the polar regions, with large patches of magnetic field of the opposite polarity situated at the north and south ``caps'' of the spherical shell, which remain largely unchanged over time (snapshots in time are not shown but available and analysed). These polar magnetic fluxes are also evident in the azimuthally averaged toroidal and poloidal field lines plotted in the meridional plane. In the vicinity of the equatorial plane up to mid latitudes, the magnetic field has a more complex quadrupolar symmetry with magnetic field patches of the
same polarity across the equatorial plane. These quadrupolar patches exhibit a dominant azimuthal wave number ($m=3$) and drift in  longitude and oscillate in latitude. Although this is, therefore, by no means a close model of the solar dynamo, we cannot help but point out
promising similarities with the latter, for example, the overall bipolar structure, rudiments of cyclicity and active longitudes, not unlike those observed on the Sun~\cite{Usoskin2007}.   {The oscillation period is directly related to the amplitude of the differential rotation (see~\cite{Busse2006}).}

\unskip

\vspace{-12pt}

\section{Conclusions}\label{555}
In this work, we have presented an initial investigation into the effects of non-uniform viscosity and entropy diffusivity on differential rotation generated by convection in rotating spherical shells. Our motivation was to address the so-called convective conundrum, in which many models of the solar dynamo do not accurately reproduce the profile of solar differential rotation as measured using helioseismology. Our approach was to focus on a small but representative region of the parameter space in order to gain a foothold, establishing a relationship between non-uniform viscosity and entropy diffusivity and differential rotation. Upon keeping all parameter values constant, with the exception of the Prandtl number, our main result is that values of the Prandtl number somewhat larger than unity ($\Pr=5$ in our case) result in a differential rotation profile whose agreement with the observed profile is much improved compared to cases with lower values of the Prandtl number and cases using only uniform profiles of the viscosity and entropy diffusivity. The results are in agreement in the equatorial region up to mid latitudes, but discrepancies remain towards the poles, where both simulations and observations are less accurate
.

For the non-uniform viscosity and entropy diffusivity profiles, the differential rotation profile is closely connected to the presence of stronger convection at the poles for higher values of $\Pr$. The
increased entropy at the poles leads to a non-trivial entropy gradient in $\theta$, which, in turn, plays a key role in the thermal wind balance, which prevents a Taylor--Proudman state. The result is that
conical rather than cylindrical profiles of differential rotation are preferred {in the presence of sufficiently large radially non-uniform profiles of viscosity and entropy diffusivity}. It is remarkable that {spherically symmetric} radial non-uniformity can lead to latitudinal non-uniformity, {which, in contrast to the simulations reported in~\cite{Miesch2006}, develops self-consistently.}

It is somewhat surprising that a large value of the Prandtl number leads to better agreement with observations. On one hand, the values of $\Pr$ in the Sun are expected to be very small based on molecular estimates~\cite{Warnecke2023}, but on the other hand, turbulent mixing suggests values of the order of unity. The agreement may, in part, be due to the region of the parameter space under consideration, which was chosen so as to capture the presumed force balance in the solar convection zone, as well as for computational feasibility. Indeed, capturing appropriate force balances may be more meaningful than trying to adhere strictly to poorly estimated and difficult-to-achieve parameter values, as recognized in similar studies of the geodynamo~\cite{Teed2023}. Our results show that solar-like differential rotation is possible in simulations, providing a good starting point for more detailed parameter sweeps---both in terms of the fundamental non-dimensional parameters and the profiles of the non-uniform viscosity and entropy diffusivity. {This will allow for a better understanding of how differential rotation affects the dynamo mechanism, which is a potential direction for further research.

\vspace{6pt}

\authorcontributions{Conceptualization, R.D.S.
; methodology, R.D.S.; software, R.D.S.; formal analysis, R.D.S. and D.T.; investigation, P.G.; resources, R.D.S.; data curation P.G.; writing---original draft preparation, R.D.S, D.M. and P.G.; writing---review and editing, R.D.S. and D.M.; visualization, P.G.; supervision, R.D.S. and D.M. All authors have read and agreed to the published version of the manuscript.}
\funding	{{This equipment was funded by BEIS capital funding via STFC capital grant ST/R00238X/1 and STFC DiRAC Operations grant ST/R001006/1. DiRAC is part of the National e-Infrastructure.} 

}

\dataavailability{{The data presented in this study are available on request from the corresponding author.}}

\acknowledgments{Numerical simulations were carried out by the DiRAC Extreme Scaling service at the University of Edinburgh, operated by the Edinburgh Parallel Computing Centre on behalf of the STFC DiRAC HPC Facility (\url{www.dirac.ac.uk}).
}

\conflictsofinterest{The authors declare no conflicts of interest.}


\begin{adjustwidth}{-\extralength}{0cm}
\reftitle{References}

\PublishersNote{}
\end{adjustwidth}

\begin{thebibliography}{999}

\bibitem[Miesch(2005)]{Miesch2005}
Miesch, M.S.
\newblock Large-Scale Dynamics of the Convection Zone and Tachocline.
\newblock {\em Living Rev. Sol. Phys.} {\bf 2005}, {\em 2}, {1}
. [\href{http://doi.org/10.12942/lrsp-2005-1}{CrossRef}]

\bibitem[Charbonneau(2020)]{Charbonneau2020}
Charbonneau, P.
\newblock Dynamo models of the solar cycle.
\newblock {\em Living Rev. Sol. Phys.} {\bf 2020}, {\em 17}, {4.} 
\newblock. [\href{http://dx.doi.org/10.1007/s41116-020-00025-6}{CrossRef}]

\bibitem[Stejko et~al.(2022)Stejko, Kosovichev, Featherstone, Guerrero,
Hindman, Matilsky, and Warnecke]{Stejko2022}
Stejko, A.M.; Kosovichev, A.G.; Featherstone, N.A.; Guerrero, G.; Hindman,
B.W.; Matilsky, L.I.; Warnecke, J.
\newblock Constraining Global Solar Models through Helioseismic Analysis.
\newblock {\em  Astrophys. J.} {\bf 2022}, {\em 934},~161. [\href{http://dx.doi.org/10.3847/1538-4357/ac7a44}{CrossRef}]

\bibitem[Simitev and Busse(2012)]{simitev2012far}
Simitev, R.D.; Busse, F.H.
\newblock How far can minimal models explain the solar cycle?
\newblock {\em  Astrophys. J.} {\bf 2012}, {\em 749},~9. [\href{http://dx.doi.org/10.1088/0004-637X/749/1/9}{CrossRef}]

\bibitem[O'Mara et~al.(2016)O'Mara, Miesch, Featherstone, and
Augustson]{OMara2016}
O'Mara, B.; Miesch, M.S.; Featherstone, N.A.; Augustson, K.C.
\newblock Velocity amplitudes in global convection simulations: The role of the
Prandtl number and near-surface driving.
\newblock {\em Adv. Space Res.} {\bf 2016}, {\em 58},~1475--1489. [\href{http://dx.doi.org/10.1016/j.asr.2016.03.038}{CrossRef}]

\bibitem[Gizon and Birch(2012)]{Gizon2012}
Gizon, L.; Birch, A.C.
\newblock Helioseismology challenges models of solar convection.
\newblock {\em Proc. Natl. Acad. Sci.} {\bf 2012},
{\em 109},~11896--11897. [\href{http://dx.doi.org/10.1073/pnas.1208875109}{CrossRef}] [\href{http://www.ncbi.nlm.nih.gov/pubmed/22814376}{PubMed}]

\bibitem[Hotta et~al.(2023)Hotta, Bekki, Gizon, Noraz, and
Rast]{hotta2023dynamics}
Hotta, H.; Bekki, Y.; Gizon, L.; Noraz, Q.; Rast, M.P.
\newblock Dynamics of solar large-scale flows. \emph{arXiv} \textbf{2023}, arXiv:2307.06481.
\newblock {\url{https://doi.org/10.48550/ARXIV.2307.06481}}.

\bibitem[Thompson et~al.(1996)Thompson, Toomre, Anderson, Antia, Berthomieu,
Burtonclay, Chitre, Christensen-Dalsgaard, Corbard, DeRosa, Genovese, Gough,
Haber, Harvey, Hill, Howe, Korzennik, Kosovichev, Leibacher, Pijpers,
Provost, Rhodes, Schou, Sekii, Stark, and Wilson]{Thompson1996}
Thompson, M.J.; Toomre, J.; Anderson, E.R.; Antia, H.M.; Berthomieu, G.;
Burtonclay, D.; Chitre, S.M.; Christensen-Dalsgaard, J.; Corbard, T.; DeRosa,
M.;  et~al.
\newblock Differential Rotation and Dynamics of the Solar Interior.
\newblock {\em Science} {\bf 1996}, {\em 272},~1300--1305. [\href{http://dx.doi.org/10.1126/science.272.5266.1300}{CrossRef}] [\href{http://www.ncbi.nlm.nih.gov/pubmed/8662459}{PubMed}]

\bibitem[Howe(2009)]{Howe2009}
Howe, R.
\newblock Solar Interior Rotation and its Variation.
\newblock {\em Living Rev. Sol. Phys.} {\bf 2009}, {\em 6}, {1}. [\href{http://dx.doi.org/10.12942/lrsp-2009-1}{CrossRef}]

\bibitem[Warnecke et~al.(2012)Warnecke, K\"{a}pyl\"{a}, Mantere, and
Brandenburg]{Warnecke2012}
Warnecke, J.; K\"{a}pyl\"{a}, P.J.; Mantere, M.J.; Brandenburg, A.
\newblock {Solar-like differential rotation and equatorward migration in a
convective dynamo with a coronal envelope}.
\newblock {\em Proc. Int. Astron. Union} {\bf 2012},
{\em 8},~307--312. [\href{http://dx.doi.org/10.1017/S1743921313002676}{CrossRef}]

\bibitem[Guerrero et~al.(2013)Guerrero, Smolarkiewicz, Kosovichev, and
Mansour]{Guerrero2013}
Guerrero, G.; Smolarkiewicz, P.K.; Kosovichev, A.G.; Mansour, N.N.
\newblock {Differential rotation in solar-loke stars from global
simulations}.
\newblock {\em  Astrophys. J.} {\bf 2013}, {\em 779},~176. [\href{http://dx.doi.org/10.1088/0004-637X/779/2/176}{CrossRef}]

\bibitem[Matilsky et~al.(2020)Matilsky, Hindman, and Toomre]{Matilsky2020}
Matilsky, L.I.; Hindman, B.W.; Toomre, J.
\newblock {Revisiting the Sun's Strong Differential Rotation along Radial
Lines}.
\newblock {\em  Astrophys. J.} {\bf 2020}, {\em 898},~111. [\href{http://dx.doi.org/10.3847/1538-4357/ab9ca0}{CrossRef}]

\bibitem[Miesch et~al.(2006)Miesch, Brun, and Toomre]{Miesch2006}
Miesch, M.S.; Brun, A.S.; Toomre, J.
\newblock {Solar Differential Rotation Influenced by Latitudinal Entropy
Variations in the Tachocline}.
\newblock {\em  Astrophys. J.} {\bf 2006}, {\em 641},~618--625. [\href{http://dx.doi.org/10.1086/499621}{CrossRef}]

\bibitem[Dikpati and Charbonneau(1999)]{Dikpati1999}
Dikpati, M.; Charbonneau, P.
\newblock A Babcock-Leighton Flux Transport Dynamo with Solar-like Differential
Rotation.
\newblock {\em  Astrophys. J.} {\bf 1999}, {\em 518},~508--520. [\href{http://dx.doi.org/10.1086/307269}{CrossRef}]

\bibitem[K{\"a}pyl{\"a} et~al.(2012)K{\"a}pyl{\"a}, Mantere, and
Brandenburg]{kapyla2012cyclic}
K{\"a}pyl{\"a}, P.J.; Mantere, M.J.; Brandenburg, A.
\newblock Cyclic magnetic activity due to turbulent convection in spherical
wedge geometry.
\newblock {\em  Astrophys. J. Lett.} {\bf 2012}, {\em 755},~L22. [\href{http://dx.doi.org/10.1088/2041-8205/755/1/L22}{CrossRef}]

\bibitem[Warnecke(2018)]{Warnecke2018}
Warnecke, J.
\newblock Dynamo cycles in global convection simulations of solar-like stars.
\newblock {\em Astron. Astrophys.} {\bf 2018}, {\em 616},~A72. [\href{http://dx.doi.org/10.1051/0004-6361/201732413}{CrossRef}]

\bibitem[Simitev et~al.(2015)Simitev, Kosovichev, and Busse]{Simitev2015}
Simitev, R.D.; Kosovichev, A.G.; Busse, F.H.
\newblock Dynamo effects near the transition from solar to anti-solar
differential rotation.
\newblock {\em  Astrophys. J.} {\bf 2015}, {\em 810},~80. [\href{http://dx.doi.org/10.1088/0004-637X/810/1/80}{CrossRef}]

\bibitem[Brun et~al.(2004)Brun, Miesch, and Toomre]{Brun2004}
Brun, A.S.; Miesch, M.S.; Toomre, J.
\newblock Global-Scale Turbulent Convection and Magnetic Dynamo Action in the
Solar Envelope.
\newblock {\em  Astrophys. J.} {\bf 2004}, {\em 614},~1073--1098. [\href{http://dx.doi.org/10.1086/423835}{CrossRef}]

\bibitem[French et~al.(2012)French, Becker, Lorenzen, Nettelmann, Bethkenhagen,
Wicht, and Redmer]{french2012ab}
French, M.; Becker, A.; Lorenzen, W.; Nettelmann, N.; Bethkenhagen, M.; Wicht,
J.; Redmer, R.
\newblock Ab initio simulations for material properties along the Jupiter
adiabat.
\newblock {\em  Astrophys. J. Suppl. Ser.} {\bf 2012}, {\em
202},~5. [\href{http://dx.doi.org/10.1088/0067-0049/202/1/5}{CrossRef}]

\bibitem[Glatzmaier and Gilman(1981)]{Glatzmaier1981}
Glatzmaier, G.A.; Gilman, P.A.
\newblock Compressible convection in a rotating spherical shell. {IV} - Effects
of viscosity, conductivity, boundary conditions, and zone depth.
\newblock {\em  Astrophys. J. Suppl. Ser.} {\bf 1981}, {\em
47},~103. [\href{http://dx.doi.org/10.1086/190753}{CrossRef}]

\bibitem[Brun and Toomre(2002)]{Brun2002}
Brun, A.S.; Toomre, J.
\newblock Turbulent Convection under the Influence of Rotation: Sustaining a
Strong Differential Rotation.
\newblock {\em  Astrophys. J.} {\bf 2002}, {\em 570},~865--885. [\href{http://dx.doi.org/10.1086/339228}{CrossRef}]

\bibitem[Brun et~al.(2022)Brun, Strugarek, Noraz, Perri, Varela, Augustson,
Charbonneau, and Toomre]{Brun2022}
Brun, A.S.; Strugarek, A.; Noraz, Q.; Perri, B.; Varela, J.; Augustson, K.;
Charbonneau, P.; Toomre, J.
\newblock Powering Stellar Magnetism: Energy Transfers in Cyclic Dynamos of
Sun-like Stars.
\newblock {\em  Astrophys. J.} {\bf 2022}, {\em 926},~21. [\href{http://dx.doi.org/10.3847/1538-4357/ac469b}{CrossRef}]

\bibitem[Warnecke et~al.(2023)Warnecke, Korpi-Lagg, Gent, and
Rheinhardt]{Warnecke2023}
Warnecke, J.; Korpi-Lagg, M.J.; Gent, F.A.; Rheinhardt, M.
\newblock Numerical evidence for a small-scale dynamo approaching solar
magnetic Prandtl numbers.
\newblock {\em Nat. Astron.} {\bf 2023}, {\em 7},~662--668. [\href{http://dx.doi.org/10.1038/s41550-023-01975-1}{CrossRef}]

\bibitem[Karak et~al.(2015)Karak, K\"{a}pyl\"{a}, K\"{a}pyl\"{a}, Brandenburg,
Olspert, and Pelt]{Karak2015}
Karak, B.B.; K\"{a}pyl\"{a}, P.J.; K\"{a}pyl\"{a}, M.J.; Brandenburg, A.;
Olspert, N.; Pelt, J.
\newblock Magnetically controlled stellar differential rotation near the
transition from solar to anti-solar profiles.
\newblock {\em Astron. Astrophys.} {\bf 2015}, {\em 576},~A26. [\href{http://dx.doi.org/10.1051/0004-6361/201424521}{CrossRef}]

\bibitem[Browning(2010)]{Browning2010}
Browning, M.K.
\newblock Differential Rotation and Magnetism in Simulations of Fully
Convective Stars.
\newblock {\em Proc. Int. Astron. Union} {\bf 2010},
{\em 6},~69--77. [\href{http://dx.doi.org/10.1017/S1743921311017467}{CrossRef}]

\bibitem[Miesch et~al.(2008)Miesch, Brun, DeRosa, and Toomre]{Miesch2008}
Miesch, M.S.; Brun, A.S.; DeRosa, M.L.; Toomre, J.
\newblock Structure and Evolution of Giant Cells in Global Models of Solar
Convection.
\newblock {\em  Astrophys. J.} {\bf 2008}, {\em 673},~557--575. [\href{http://dx.doi.org/10.1086/523838}{CrossRef}]

\bibitem[Sasaki et~al.(2018)Sasaki, ichi Takehiro, Ishiwatari, and
Yamada]{Sasaki2018}
Sasaki, Y.; ichi Takehiro, S.; Ishiwatari, M.; Yamada, M.
\newblock Effects of radial distribution of entropy diffusivity on critical
modes of anelastic thermal convection in rotating spherical shells.
\newblock {\em Phys. Earth Planet. Inter.} {\bf 2018}, {\em
276},~36--43. [\href{http://dx.doi.org/10.1016/j.pepi.2017.09.003}{CrossRef}]

\bibitem[Jones et~al.(2011)Jones, Boronski, Brun, Glatzmaier, Gastine, Miesch,
and Wicht]{Jones2011}
Jones, C.; Boronski, P.; Brun, A.; Glatzmaier, G.; Gastine, T.; Miesch, M.;
Wicht, J.
\newblock Anelastic convection-driven dynamo benchmarks.
\newblock {\em Icarus} {\bf 2011}, {\em 216},~120--135. [\href{http://dx.doi.org/10.1016/j.icarus.2011.08.014}{CrossRef}]

\bibitem[Tilgner(1999)]{Tilgner1999}
Tilgner, A.
\newblock Spectral methods for the simulation of incompressible flows in
spherical shells.
\newblock {\em Int. J. Numer. Methods Fluids} {\bf
1999}, {\em 30},~713--724.
\newblock
[\href{http://dx.doi.org/10.1002/(SICI)1097-0363(19990730)30:6<713::AID-FLD859>3.0.CO;2-Y}{CrossRef}]

\bibitem[Tassoul(2000)]{Tassoul2000}
Tassoul, J.L.
\newblock {\em Stellar Rotation}; Cambridge University Press: {Cambridge, UK,} 2000. [\href{http://dx.doi.org/10.1017/cbo9780511546044}{CrossRef}]

\bibitem[Simitev and Busse(2003)]{Simitev2003}
Simitev, R.; Busse, F.H.
\newblock Patterns of convection in rotating spherical shells.
\newblock {\em New J. Phys.} {\bf 2003}, {\em 5},~97--97. [\href{http://dx.doi.org/10.1088/1367-2630/5/1/397}{CrossRef}]

\bibitem[Simitev and Busse(2005)]{SIMITEV2005}
Simitev, R.D.; Busse, F.H.
\newblock Prandtl-number dependence of convection-driven dynamos in rotating
spherical fluid shells.
\newblock {\em J. Fluid Mech.} {\bf 2005}, {\em 532},~365--388. [\href{http://dx.doi.org/10.1017/S0022112005004398}{CrossRef}]

\bibitem[K\"{a}pyl\"{a}(2023)]{Kapyla2023}
K\"{a}pyl\"{a}, P.J.
\newblock Transition from anti-solar to solar-like differential rotation:
Dependence on Prandtl number.
\newblock {\em Astron. Astrophys.} {\bf 2023}, {\em 669},~A98. [\href{http://dx.doi.org/10.1051/0004-6361/202244395}{CrossRef}]

\bibitem[Ardes et~al.(1997)Ardes, Busse, and Wicht]{Ardes1997}
Ardes, M.; Busse, F.; Wicht, J.
\newblock Thermal convection in rotating spherical shells.
\newblock {\em Phys. Earth Planet. Inter.} {\bf 1997}, {\em
99},~55--67. [\href{http://dx.doi.org/10.1016/S0031-9201(96)03200-1}{CrossRef}]

\bibitem[BUSSE and SIMITEV(2004)]{BUSSE2004}
Busse, F.H.; Simitev, R.
\newblock Inertial convection in rotating fluid spheres.
\newblock {\em J. Fluid Mech.} {\bf 2004}, {\em 498},~23--30. [\href{http://dx.doi.org/10.1017/S0022112003006943}{CrossRef}]

\bibitem[Zhang and Busse(1987)]{Zhang1987}
Zhang, K.K.; Busse, F.H.
\newblock On the onset of convection in rotating spherical shells.
\newblock {\em Geophys. Astrophys. Fluid Dyn.} {\bf 1987}, {\em
39},~119--147. [\href{http://dx.doi.org/10.1080/03091928708208809}{CrossRef}]

\bibitem[Kosovichev(1996)]{Kosovichev1996}
Kosovichev, A.G.
\newblock Helioseismic Constraints on the Gradient of Angular Velocity at the
Base of the Solar Convection Zone.
\newblock {\em  Astrophys. J.} {\bf 1996}, {\em 469},~L61--L64. [\href{http://dx.doi.org/10.1086/310253}{CrossRef}]

\bibitem[Pedlosky(1987)]{Pedlosky1987}
Pedlosky, J.
\newblock {\em Geophysical Fluid Dynamics}; Springer: New York, NY, USA, 1987. [\href{http://dx.doi.org/10.1007/978-1-4612-4650-3}{CrossRef}]

\bibitem[Busse and Simitev(2005)]{Busse2005}
Busse, F.H.; Simitev, R.
\newblock Dynamos driven by convection in rotating spherical shells.
\newblock {\em Astron. Nachrichten} {\bf 2005}, {\em 326},~231--240. [\href{http://dx.doi.org/10.1002/asna.200410382}{CrossRef}]

\bibitem[Mather and Simitev(2020)]{Mather2020}
Mather, J.F.; Simitev, R.D.
\newblock Regimes of thermo-compositional convection and related dynamos in
rotating spherical shells.
\newblock {\em Geophys. Astrophys. Fluid Dyn.} {\bf 2020}, {\em
115},~61--84. [\href{http://dx.doi.org/10.1080/03091929.2020.1762875}{CrossRef}]

\bibitem[Gilman and Foukal(1979)]{Gilman1979}
Gilman, P.A.; Foukal, P.V.
\newblock Angular velocity gradients in the solar convection zone.
\newblock {\em  Astrophys. J.} {\bf 1979}, {\em 229},~1179. [\href{http://dx.doi.org/10.1086/157052}{CrossRef}]

\bibitem[Usoskin et~al.(2007)Usoskin, Berdyugina, Moss, and
Sokoloff]{Usoskin2007}
Usoskin, I.; Berdyugina, S.; Moss, D.; Sokoloff, D.
\newblock Long-term persistence of solar active longitudes and its implications
for the solar dynamo theory.
\newblock {\em Adv. Space Res.} {\bf 2007}, {\em 40},~951--958. [\href{http://dx.doi.org/10.1016/j.asr.2006.12.050}{CrossRef}]

\bibitem[Busse and Simitev(2006)]{Busse2006}
Busse, F.H.; Simitev, R.D.
\newblock {Parameter dependences of convection-driven dynamos in rotating
spherical fluid shells}.
\newblock {\em Geophys. Astrophys. Fluid Dyn.} {\bf 2006}, {\em
100},~341--361. [\href{http://dx.doi.org/10.1080/03091920600784873}{CrossRef}]

\bibitem[Teed and Dormy(2023)]{Teed2023}
Teed, R.J.; Dormy, E.
\newblock Solenoidal force balances in numerical dynamos.
\newblock {\em J. Fluid Mech.} {\bf 2023}, {\em 964}, {A26}. [\href{http://dx.doi.org/10.1017/jfm.2023.332}{CrossRef}]

\end{thebibliography}
\end{document}